\documentclass[11pt,A4paper]{article}
 \usepackage{graphicx}
 \usepackage{amsmath}
 \usepackage{amssymb}
 \usepackage{epsfig}
 \usepackage{natbib}    
 \usepackage[utf8]{inputenc}
 \usepackage{authblk}

 \usepackage{setspace}
 \def\blfootnote{\xdef\@thefnmark{}\@footnotetext} 
 \long\def\symbolfootnote[#1]#2{\begingroup\def\thefootnote{\fnsymbol{footnote}}\footnote[#1]{#2}\endgroup} 
\title{Generation of scenarios from calibrated ensemble forecasts \\ with a dual ensemble copula coupling approach }

\author[a,b]{Zied Ben Bouall\`egue}
\author[a]{Tobias Heppelmann}
\author[a]{Susanne E. Theis}
\author[c]{Pierre Pinson}
\affil[a]{Deutscher Wetterdienst, Offenbach, Germany}
\affil[b]{Meteorological Institute, University of Bonn, Germany}
\affil[c]{Technical University of Denmark, Denmark}
\date{}

\begin{document}
 
 \maketitle

 \begin{abstract}
Probabilistic forecasts in the form of ensemble of scenarios are required for complex decision making processes.  Ensemble forecasting systems provide such products but the spatio-temporal structures of the forecast uncertainty is lost when statistical calibration of the ensemble forecasts is applied for each lead time and location independently. Non-parametric approaches allow the reconstruction of spatio-temporal joint probability distributions at a low computational cost. For example, the ensemble copula coupling (ECC) method rebuilds the multivariate aspect of the forecast from the original ensemble forecasts. Based on the assumption of error stationarity, parametric methods aim to fully describe the forecast dependence structures. In this study, the concept of ECC is combined with past data statistics in order to account for the autocorrelation of the forecast error. 
The new approach, called d-ECC, is applied to wind forecasts from the high resolution ensemble system COSMO-DE-EPS run operationally at the German weather service. 
Scenarios generated by ECC and d-ECC are compared and  assessed in the form of time series  by means of multivariate verification tools and in a product oriented framework. Verification results over a 3 month period show that the innovative method d-ECC outperforms or performs as well as ECC in all investigated aspects. 
\end{abstract}

\section{Introduction}

Uncertainty information is essential for an optimal use of a forecast \citep{krz83}. Such information can be provided by an Ensemble Prediction System (EPS) which aims at describing the flow-dependent forecast uncertainty \citep{leut08}. Several deterministic forecasts  are run simultaneously accounting for uncertainties in the description of the initial state, the model parametrization and, for limited area models, the boundary conditions. Probabilistic products are derived from an ensemble, tailored to specific user's need. For example, wind forecasts in the form of quantiles at selected probability levels are of particular interest for actors in the renewable energy sector \citep{pinson2013}.

However, probabilistic products generally suffer from a lack of reliability, the system showing biases and failing to fully represent the forecast uncertainty. Statistical techniques allow to adjust the ensemble forecast correcting for systematic inconsistencies  \citep{gneit2007}. This step known as calibration is based on past data and usually focuses on a single or few  aspects of the ensemble forecast. For example, calibration of wind forecast can be performed by univariate approaches \citep{bremnes2004,sloughter2010,thorar2010} or bivariate methods which account for correlation structures of the wind components \citep{pinson2012,schuhen2012}. These calibration procedures provide reliable predictive probability distribution of wind speed or wind components for each forecast lead time and location independently. 
Decision making problems can however require information about the spatial and/or temporal structure of the forecast uncertainty. Examples of application in the renewable energy sector resemble  the optimal operation of a wind-storage system in a market environment, the  unit commitment over a control zone or the  optimal maintenance planning \citep{pins09}.  
In other words, scenarios that describe spatio-temporal wind variability are relevant products for end-users of wind forecasts.

The generation of scenarios from calibrated ensemble forecasts is a step that can be performed with the use of empirical copulas. The empirical copula approaches are non-parametric and, in comparison with parametric approaches \citep{keune2014,feldmann2015}, simple to implement and computationally cheap. Empirical copulas can be based on climatological records \citep[Schaake Shuffle (ScSh);][]{clark2004} or on the original raw ensemble \citep[ensemble copula coupling (ECC);][]{schefzik2014}. 
ECC, which consists in the conservation of the ensemble member rank structure from the original  ensemble to the calibrated one, has the advantage to be applicable to any location of the model domain without restriction related to the availability of observations. 
However, unrealistic scenarios  can be generated by the ECC approach 
when the post-processing  indiscriminately increases the ensemble spread to a large extent. Non-representative correlation structures in the raw ensemble are magnified after calibration leading to unrealistic forecast variability. As a consequence, ECC can deteriorate the ensemble information content when applied to ensembles with relatively poor reliability as suggested, for example, by verification results in \cite{flowerdew2014}. 

In this paper, a new version of the ECC approach is proposed in order to overcome the generation of unrealistic scenarios. 
Focusing on time series, a temporal component is introduced in the ECC scheme accounting for the autocorrelation of the forecast error over consecutive forecast lead times. The assumption of forecast error stationarity,  already adopted for the development of fully parametric approaches  \citep{pins09,schoelzel2011}, is exploited in combination with the structure information of the original scenarios.
The new approach  based on these two sources of information, past data and ensemble structure,  
is called \textit{dual} ensemble copula coupling (d-ECC).
Objective verification  is performed in order to show the benefit of the proposed approach with regard to the standard ECC.

The manuscript is organized as follows: Section \ref{sec:data} describes the dataset used to illustrate the manuscript as well as the calibration method applied to derive calibrated quantile forecasts from the raw ensemble. Sections \ref{sec:method} and \ref{sec:illustration} introduce the empirical copula approaches  for the generation of scenarios and discuss in particular the ECC and d-ECC methods. Section \ref{sec:verifmethod} describes the verification process for the scenario assessment. Section \ref{sec:results} presents the results obtained by means of multivariate scores and in a product oriented verification framework.

\section{Data}

\label{sec:data}
\subsection{Ensemble forecasts and observations}   
COSMO-DE-EPS is the high resolution ensemble prediction system run operationally at DWD. It consists of 20 COSMO-DE forecasts with variations in the initial conditions, the boundary conditions and the model physics \citep{gtpb10,pbtg12}. COSMO-DE-EPS follows the multi-model ensemble approach, with 4 global models driving each 5 physically perturbed members. The ensemble configuration implies a clustering of the ensemble members as a function of the driving global model when large scale structures dominate the forecast uncertainty. 

\begin{figure}
\centerline{\includegraphics[width=0.4\textwidth]{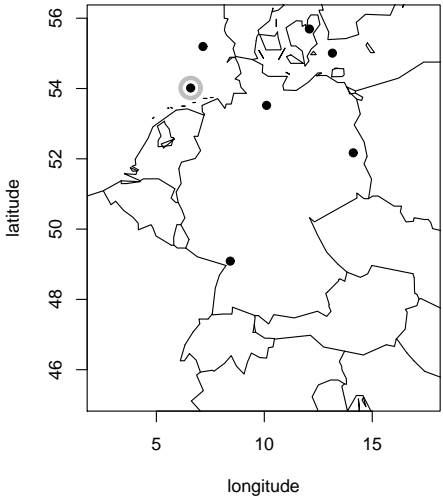}}
\caption{
  Map of Germany and neighboring areas (approximately the COSMO-DE domain) with latitude/longitude axes. Location of the 7 wind stations used in this study. The station FINO1 is highlighted with a grey mark.
}
 \label{fig:map}
\end{figure}

The focus is here on wind forecasts at 100 meter height above ground. The post-processing methods are applied to forecasts of the 00UTC run with an hourly output interval and a forecast horizon of up to 21 hours. The observation dataset comprises quality controlled wind measurements from 7 stations: Risoe, FINO1, FINO2, FINO3, Karlsruhe, Hamburg and Lindenberg, as plotted in Figure \ref{fig:map}.  The verification period covers a 3 month period: March, April and May 2013.

\begin{figure*}
\centerline{\includegraphics[width=\textwidth]{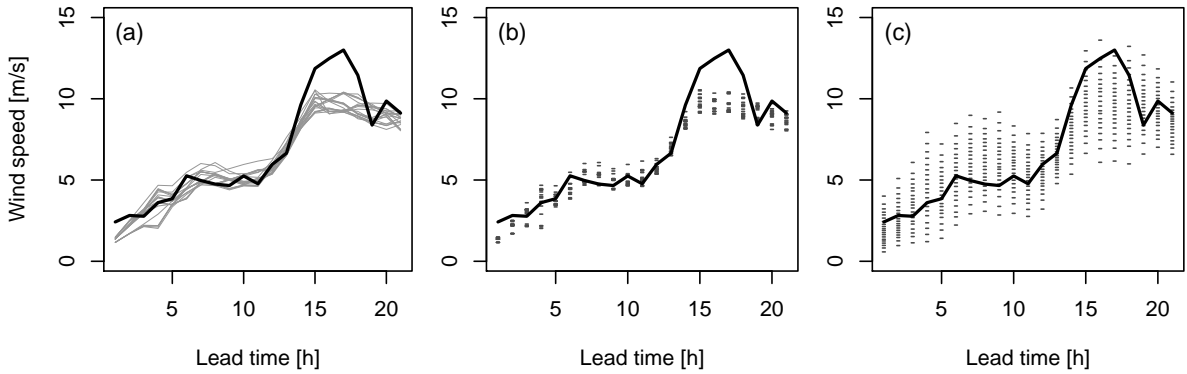}}
\caption{
    Wind speed at 100 meter height above ground, on March 2, 2013, at station FINO1: (a) COSMO-DE-EPS forecast (grey lines), (b) raw ensemble forecast in the form of quantiles (sorted members, see text), (c) calibrated quantile forecasts, and the corresponding observations (black lines).
}
\label{fig:ex}
\end{figure*}

Figure \ref{fig:ex}(a) shows an example of a COSMO-DE-EPS wind forecast at  hub-height. The forecast is valid on day March 2, 2013, at station FINO1 (see Figure \ref{fig:map}). The ensemble members are drawn in grey while the corresponding observations are drawn in black. In Figure \ref{fig:ex}(b), the raw ensemble forecast is interpreted in the form of quantiles. 

Formally, a quantile  \(q_{\tau}\) at probability level \(\tau\) (with \(\ 0 \le \tau \le 1\)) is defined as: 
\begin{equation}
q_{\tau} := F^{-1}(\tau)=\text{inf}\{y:F(y)\ge \tau \}
\label{equ:cdf}
\end{equation}
where $F$ is the cumulative probability distribution of the random variable \(Y \in \Re\):
\begin{equation}
 F(y)=\mathbb{P}(Y \le y ).
\label{equ:cdf}
\end{equation}
In practice, at each forecast lead time, the member of rank $n$ can be interpreted as a quantile forecast at probability level $\tau_n$:
\begin{equation}
   \tau_n= \frac{n}{N_e+1} 
\label{equ:tau}
\end{equation}
where $N_e$ is the number of ensemble members.

In the example of Figure \ref{fig:ex}, the raw ensemble is not able to capture the observation variability. Calibration aims to correct for this lack of reliability by adjusting the mean and enlarging the spread of the ensemble forecast.

\subsection{Calibrated ensemble forecasts}   
Since COSMO-DE-EPS forecasts have shown to suffer from statistical inconsistencies \citep{zbb12,zbbmausam}, calibration has to be applied in order to provide reliable forecasts to the users.  
The method applied in this study is the bivariate Non-homogeneous Gaussian Regression \citep[EMOS,][]{schuhen2012}. The mean and variance of each wind component as well as the correlation between the two components characterize the predictive bivariate normal distribution. Corrections applied to the raw ensemble mean and variance are optimized by minimizing the continuous ranked probability score \citep[$CRPS$;][]{matheson76}. The calibration coefficients are estimated for each station and each lead time separately (local version of EMOS), based on a training period being defined as a moving window of 45 days. 

The final calibrated products considered here are $N_e$ equidistant 
 forecasts of wind speed estimated for each location and each forecast lead time separately, where the $N_e$ probability levels associated to the forecast quantiles follow Eq.~\eqref{equ:tau}. 
Calibrated quantile forecasts are shown in Figure \ref{fig:ex}(c). The spread of the ensemble is increased with respect to Figure \ref{fig:ex}(b) and thus the observation variability is now captured by the forecast. From a statistical point of view the calibration method provides reliable ensemble marginal distributions and reliable quantile forecasts as checked by means of rank histograms and quantile reliability plots (not shown). The performance of the applied calibration technique  is similar to the one obtained by other methods such as quantile regression \citep{koenker78,bremnes2004}. 

Information about spatial and temporal dependence structures, which are crucial in many applications, are however not available any more after this calibration step (see Figure \ref{fig:ex}(c)). The next post-processing step consists then in the generation of consistent scenarios based on the calibrated samples.

\section{Generation of scenarios }
\label{sec:method}

The generation of scenarios with empirical copulas is here briefly described. For a deeper insight into the methods, the reader is invited to refer to the original article of \cite{schefzik2014},  or to \cite{wilks2014} and references within. 

First, consider the multivariate cumulative distribution function (\textit{cdf}) $G$ defined as:
\begin{equation}
G(y_1,...,y_L) = \mathbb{P}[Y_1\le y_1,...,Y_L\le y_L]
\label{equ:sklar}
\end{equation}
of a random vector $(Y_1,...,Y_L)$ with $y_1,...,y_L \in \mathbb{R}$.
As in Eq.~\eqref{equ:cdf}, we define the marginals $F_i$ as:
\begin{equation}
F_i(y_i) = \mathbb{P}[Y_i\le y_i].
\label{equ:sklar}
\end{equation}
The Sklar's theorem \citep{sklar59} states that G can be expressed as:
\begin{equation}
G(y_1,...,y_L) = C(F_1(y_1),F_L(y_L))
\label{equ:sklar}
\end{equation}
where $C$ is a copula that links an L-variate cumulative distribution function  $G$ to its univariate marginal \textit{cdf}s $F_1,...,F_L$.

In Eq.~\eqref{equ:sklar}, a joint distribution is represented as univariate margins plus copulas. The problem of estimating univariate distributions and the problem of estimating dependence can therefore be treated separately.  Univariate calibration marginal \textit{cdf}s $F_1,...,F_L$ are provided by the  calibration step described in the previous section. 
The choice of the copula $C$ depends on the application and on the size $L$ of the multivariate problem. We focus here on empirical copulas since they are suitable for problems with high dimensionality.  

We denote $H$ the empirical copula.  $H$ is based on a multivariate dependence template, a specific discrete dataset $\boldsymbol{z}$ defined in $ \mathbb{R}^L$. 
The chosen dataset is described formally as:
\begin{equation}
\boldsymbol{z}:=  \left\{(z_1^1,...,z_1^N),...,(z_L^1,...,z_L^N) \right\}
\label{equ:defz}
\end{equation}
consisting of $L$ tuples of size $N$ with entries in $\mathbb{R}$. In other words, $L$ is the dimension of the multivariate variable and $N$ is the number of scenarios. 
The rank of $z_l^n$ for $n\in\left\{1,...,N \right\} $ and  $l\in\left\{1,...,L\right\} $ is defined as:
\begin{equation}
R_l^n:= \sum_{i=1}^N \mathbb{I}(z_l^i \le z_l^n) 
\end{equation}
where $\mathbb{I}(\cdot)$ denotes the indicator function taking value 1 if the condition in parenthesis is true and zero otherwise. 
The empirical copula $H$ induced by the dataset $\boldsymbol{z}$ is given by:
\begin{align}
H(\frac{j_1}{N},...,\frac{j_L}{N}) &:= \frac{1}{N}\sum_{i=1}^{N} \mathbb{I}(R_1^i \le j_1,...,R_L^i \leq j_L ) \\
&= \frac{1}{N}\sum_{i=1}^{N}  \prod_{l=1}^{L}  \mathbb{I}(R_l^i \le j_l)  
\end{align}
for integers $0 \le j_1,...,j_L \le N $.

In practice, $N$ equidistant quantiles of $F_l$ with  $l\in\left\{1,...,L \right\}$ are derived from the univariate calibration step:
\begin{equation}
\boldsymbol{q}:= \left\{(q_1^1,...,q_1^N),...,(q_L^1,...,q_L^N) \right\}
\label{equ:qens}
\end{equation}
with 
\begin{equation}
q_l^n:= F_l^{-1} \left(\tau_n \right);\quad n \in\left\{ 1,..,N\right\} 
\label{equ:qmarg}
\end{equation}
where $\tau_n$ is defined in Eq.  \eqref{equ:tau}. The sample  $\boldsymbol{q}$ is rearranged following the dependence structure of the reference template $\boldsymbol{z}$. 
The permutations $\pi_{l}(n):=R_l^n$ for $ n \in\left\{ 1,..,N\right\}$ 
are derived  from 
the univariate  ranks  $R_l^{1}, ..., R_l^{N}$ 
for $ l \in\left\{ 1,..,L\right\}$ and applied to the univariate calibrated sample $\boldsymbol{q}$. The post-processed scenarios $\tilde{x}_l^1,...,\tilde{x}_l^N$ for each margin $l$ is expressed as:
\begin{equation}
\tilde{x}_l^1:= q_l^{\pi_l(1)},..., \tilde{x}_l^N:= q_l^{\pi_l(N)}
\label{equ:pp}
\end{equation}

The multivariate correlation structures are generated based on the rank correlation structures of a sample template $\boldsymbol{z}$. The empirical copulas  presented here only differ in the way $\boldsymbol{z}$ is defined. In the following, let $t \in \left\{1,\dots,T\right\}$ be a lead time and let $L:=T$. 
For simplicity, we consider here a single weather variable and a single location.

\subsection{Ensemble copula coupling}
The rank structure of the ensemble is preserved after calibration when applying the standard ensemble copula coupling approach (ECC). The raw ensemble forecast is denoted $\boldsymbol{x}$:

\begin{equation}
\boldsymbol{x}:= \left\{(x_1^1,...,x_1^{N_e}),...,(x_L^1,...,x_L^{N_e}) \right\}
\label{equ:rawe}
\end{equation}
where $N_e$ is the ensemble size. ECC applies without restriction to any multivariate setting. The number of scenarios generated with ECC is however the same as the size of the original ensemble ($N=N_e$).  The transfer of the rank structure from the raw ensemble forecast to the calibrated one consists then in taking $\boldsymbol{x}$ as the required template in Eq.~\eqref{equ:defz}.  

\begin{figure*}
\centerline{\includegraphics[width=\textwidth]{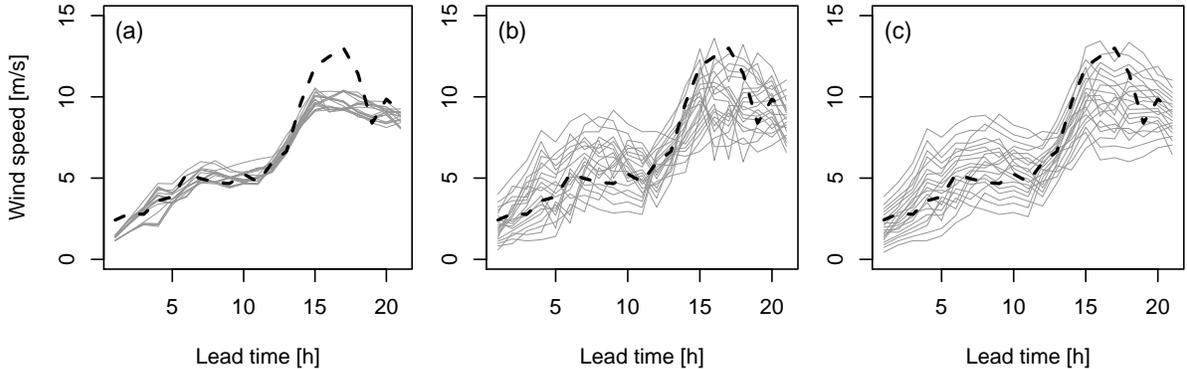}}
\caption{
   Same example as in Figure \ref{fig:ex}:  (a) COSMO-DE-EPS scenarios, (b) ECC derived scenarios, (c) d-ECC derived scenarios, and the corresponding observations (black lines).
}
\label{fig:exSC}
\end{figure*}

Based on COSMO-DE-EPS forecasts of Figure \ref{fig:exSC}(a) (identical to Figure  \ref{fig:ex}(a)), an example of scenarios derived with ECC  is provided in Figure \ref{fig:exSC}(b). 
The increase of spread after the calibration step implies a larger step-to-step variability in the time trajectories. Figure \ref{fig:excECC} focuses on a single scenario highlighting the difference between the original and post-processed scenarios.

\subsection{Dual ensemble copula coupling}
ECC assumes that the ensemble prediction system correctly describes the spatio-temporal dependence structures of the weather variable. This assumption is quite strong and cannot be valid in all cases.  On the other side, based on the assumption of error stationarity, parametric methods have been developed  focusing on covariance structures of the forecast error \citep{pins09,schoelzel2011}. We propose a new version of the ECC approach which is an attempt to combine  both information: the structure of the original ensemble and the error autocorrelation estimated from past data. Therefore, the new scheme is called dual ensemble copula coupling (d-ECC) as the copula relies on a dual source of information.

For this purpose, we denote $\boldsymbol{e}$ the forecast error defined as the difference between ensemble mean forecasts and observations:
\begin{align}
\boldsymbol{e} &:= \left\{e_1,...,e_T\right\} \\
&= \left\{ {y}_1 - m({x}_1),...,{y}_T  - m({x}_T) \right\} 
\label{equ:d}
\end{align}
where $m({x}_t)$  and  ${y}_{t}$ are the ensemble mean and the corresponding observation at lead time $t \in \left\{1,...,T\right\}$, respectively. The temporal correlation of the error is described by a correlation matrix $\boldsymbol{R_e}$ defined as:
\begin{equation}
\boldsymbol{R_e} =  
\begin{pmatrix}
  \rho_{e_1,e_1} & \rho_{e_1,e_2} & \cdots & \rho_{e_1,e_T} \\
  \rho_{e_2,e_1} & \rho_{e_2,e_2} & \cdots & \rho_{e_2,e_T} \\
  \vdots  & \vdots  & \ddots & \vdots  \\
  \rho_{e_T,e_1} & \rho_{e_T,e_2} & \cdots & \rho_{e_T,e_T} 
 \end{pmatrix}
\label{equ:cormat}
\end{equation}
where $\rho_{e_{t_1},e_{t_2}}$ is the correlation coefficient of the forecast error at lead times $t_1$ and $t_2$. 
The  empirical correlation matrix  $\boldsymbol{\hat{R}_e}$   is estimated   
based on the training samples used for the univariate calibration step at the different lead times. In our setup,  $\boldsymbol{\hat{R}_e}$ is regularly updated on a daily basis from the moving windows of 45 days defined as training datasets for the EMOS application.

Again here, we aim at constructing a template (Eq.~\ref{equ:defz}) in order to establish  the correlation structures within the calibrated ensemble $\boldsymbol{q}:= \left\{(q_1^1,...,q_1^{N_e}),...,(q_T^1,...,q_T^{N_e}) \right\}$. 
In the d-ECC approach, the template is built performing the following steps:
\begin{enumerate}
\item \label{item:ECC}
Apply ECC with  the original ensemble forecast $\boldsymbol{x}$ as reference sample template, in order to derive a post-processed ensemble of scenarios  $\boldsymbol{\tilde{x}}$:
\begin{equation}
\boldsymbol{\tilde{x}}:= \left\{(\tilde{x}_1^1,...,\tilde{x}_1^{N_e}),...,(\tilde{x}_T^1,...,\tilde{x}_T^{N_e}) \right\},
\label{equ:cens}
\end{equation}
\item \label{item:bias} Derive the error correction  $\boldsymbol{c}^i$  imposed to each scenario $i$  ($i \in 1,...,N_e$) of the reference template by this post-processing step:
\begin{align}
\boldsymbol{c}^i &:= \left\{c^i_1,...,c^i_T\right\} \\
&= \left\{ \tilde{x}^i_1 - x^i_1,...,\tilde{x}^i_T -x^i_T \right\},
\label{equ:d}
\end{align}

\item \label{item:biascor} \textit{Transformation step}: Apply a transformation to the correction $\boldsymbol{c}^i$  of each scenario based on the estimate of the error autocorrelation $\boldsymbol{\hat{R}_e}$ and its eigendecomposition $\boldsymbol{\hat{R}_e}=\boldsymbol{U}\boldsymbol{\Lambda}\boldsymbol{U^{-1}}$ in order to derive the \textit{adjusted corrections } $\boldsymbol{\breve{c}}^i$:
\begin{align}
\boldsymbol{\breve{c}}^i &= \boldsymbol{\hat{R}_e}^{\frac{1}{2}} {\boldsymbol{c}^{i}} \\
&= \boldsymbol{U} \boldsymbol{\Lambda}^{\frac{1}{2}} \boldsymbol{U^{-1}} \boldsymbol{c}^{i},
\label{eq:biascor}
\end{align}

\item Derive the  so-called \textit{adjusted ensemble}  $\boldsymbol{\breve{x}}$:
\begin{equation}
\boldsymbol{\breve{x}}:= \left\{(\breve{x}_1^1,...,\breve{x}_1^{N_e}),...,(\breve{x}_T^1,...,\breve{x}_T^{N_e}) \right\}
\label{equ:aens}
\end{equation}
where a scenario $\boldsymbol{\breve{x}}^i= \left\{\breve{x}_1^i,...,\breve{x}_T^i) \right\}$  of $\boldsymbol{\breve{x}}$ is defined as a combination of the original member and the adjusted error correction:
\begin{equation}
\boldsymbol{\breve{x}}^i = \boldsymbol{x}^i+ \boldsymbol{\breve{c}}^i,
\label{equ:kecc2}
\end{equation}

\item 
Take $\boldsymbol{\breve{x}}$ as reference template in Eq.~\eqref{equ:defz} so that the new empirical copula is based on the adjusted ensemble.

\end{enumerate}
The d-ECC reference template $\boldsymbol{\breve{x}}$ combines the raw ensemble structure and the autocorrelation of the forecast error reflected in the adjusted member corrections. The transformation of the scenario corrections in Eq.~\eqref{eq:biascor} adjusts their correlation structure based on the error correlation matrix $\boldsymbol{\hat{R}_e}$. Taking the square root of the correlation matrix (Eq.~\ref{eq:biascor}) resembles a signal processing technique which is described as a \textit{coloring transformation} of a vector of random variables \citep{kessy2015}.

\section{Illustration and discussion of d-ECC}
\label{sec:illustration}

\begin{figure}
\centerline{\includegraphics[width=7cm]{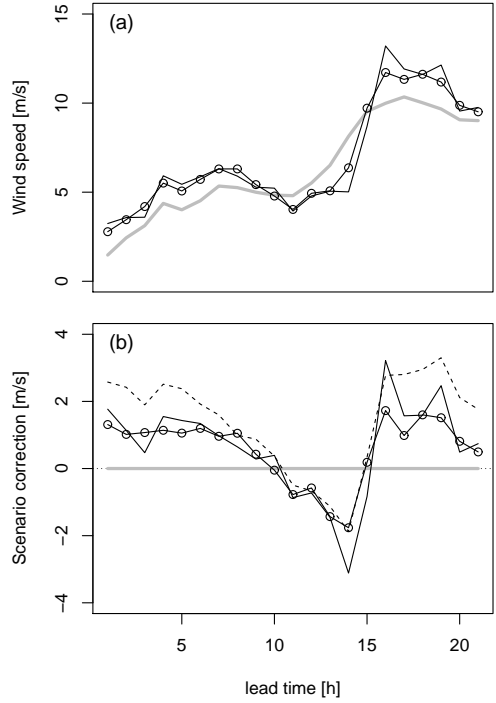}}
\caption{
    Illustration of the concept of d-ECC based on the example of Figure \ref{fig:exSC} showing (a) one among the 20 scenarios and (b) the correction applied to the original scenario after post-processing. The raw ensemble forecast (here the  member 13)  is represented  in grey, the ECC scenario in black, and the  d-ECC scenario in black with dots. The dashed line represents the scenario correction adjusted by the transformation step (see text).  
}
\label{fig:excECC}
\end{figure}

Focusing on a single member, 
the d-ECC steps are illustrated in Figure \ref{fig:excECC}. First,  the  correction associated to each ECC scenario with respect to the corresponding original ensemble member is computed (black line in Figure \ref{fig:excECC}(b), Eq.~\ref{equ:d}). This scenario correction is adjusted based on the assumption of temporal autocorrelation of the error (dashed line in Figure \ref{fig:excECC}(b), Eq.~\ref{eq:biascor}). This adjusted scenario correction is then superimposed on the original ensemble forecast before to draw again the correlation structure of the adjusted ensemble. 

The new scheme reduces to the standard ECC in the  case where $rank({x}_t^i) =  rank(\breve{x}_t^i)$ for all $i \in  \left\{1,...,N_e\right\}$ and $t \in  \left\{1,\dots,T\right\}$, which means that the additional terms $\boldsymbol{\breve{c}}^i$ do not have any impact on the rank structure of the ensemble.  This case occurs if:
\begin{itemize}
\item $\boldsymbol{\hat{R}_e}= \boldsymbol{I}$ where $\boldsymbol{I}$ is the identity matrix,  which means that there is no temporal correlation of the error in the original ensemble,
\item  $\boldsymbol{c} = \boldsymbol{0}$  where $\boldsymbol{0}$ is the null vector, which means that the calibration step does not impact the forecast, the forecast being already well calibrated.
\item  $\boldsymbol{c} = h\cdot \boldsymbol{J}$  where $h$ is a constant and $\boldsymbol{J}$ an all-ones vector, which means that  the calibration step corrects only for bias errors and the system is spread bias free.
\end{itemize}
So the d-ECC typically takes effect if calibration corrects the spread and if this correction is correlated in time at the member level. 

Some more insight can be gained by looking at the following equations. Let the observation $y_t$ and the postprocessed ensemble members $\tilde{x}^i_t$ be realizations of random variables $Y$ and $\tilde{X}$.
Consider the covariance of the forecast error denoted $k$ and defined as: 
\begin{equation}
	k_{t_{1},t_{2}}:=\mathbb{E}[(Y_{t_1} - m(\tilde{X}_{t_1}))(Y_{t_2}- m(\tilde{X}_{t_2}))]
\label{equ:dcdef1}
\end{equation}
where $t_1$ and $t_2$ are two lead times and $\mathbb{E}[\cdot]$ the expectation operator. It is assumed that the postprocessed ensemble mean $m(\tilde{x}_t)$ is fully bias-corrected so that \mbox{$\mathbb{E}[Y_{t} - m(\tilde{X}_{t})]=0$}.

After post-processing, the forecast scenarios and observation time series are considered as drawn from the same multivariate probability distribution, so the forecast error covariance can also be expressed as:  
\begin{align}
k_{t_{1},t_{2}} &=\mathbb{E}[(\tilde{X}_{t_1} - m(\tilde{X}_{t_1}))(\tilde{X}_{t_2}- m(\tilde{X}_{t_2}))]  \\ 
&=  \rho_{\tilde{x}_{t_1},\tilde{x}_{t_2}} \sigma_{\tilde{x}_{t_1}} \sigma_{\tilde{x}_{t_2}}
\label{equ:dcdef2}
\end{align}
where $\rho_{\tilde{x}_{t_1},\tilde{x}_{t_2}}$ refers to the correlation between $\tilde{x}_{t_1}$ and $\tilde{x}_{t_2}$ and $\sigma_{\tilde{x}_{t}}$ refers to the square root of the variances between
the members of the calibrated ensemble $(\tilde{x}^1,...,\tilde{x}^{N_e})$ at lead time $t$.
The corresponding estimators are the following:
\begin{equation}
\hat{k}_{t_1,t_2}=\frac{1}{N_e-1}\sum_{i=1}^{N_e}[(\tilde{x}^i_{t_1} - m(\tilde{x}_{t_1}))(\tilde{x}^i _{t_2}- m(\tilde{x}_{t_2}))]
\label{equ:dcdef}
\end{equation}
and
\begin{equation}
\hat{\sigma}_{\tilde{x}_t} = \sqrt{\frac{1}{N_e-1}\sum_{i=1}^{N_e}(\tilde{x}^i_t - m(\tilde{x}_t))^2}
\end{equation}
and 
\begin{equation}
\hat{\rho}_{\tilde{x}_{t_1},\tilde{x}_{t_2}}=\frac{\hat{k}_{t_1,t_2}}{\hat{\sigma}_{\tilde{x}_{t_1}} \hat{\sigma}_{\tilde{x}_{t_2}}}.
\end{equation}
From Eq.~\eqref{equ:d} recall that
\begin{equation}
\tilde{x}^i_t = x^i_t + c^i_t 
\end{equation}
so we can rewrite the expression in Eq.~\eqref{equ:dcdef2} as
\begin{equation}
\rho_{\tilde{x}_{t_1},\tilde{x}_{t_2}} \sigma_{\tilde{x}_{t_1}} \sigma_{\tilde{x}_{t_2}} = {\rho}_{x_{t_1},x_{t_2}}  \sigma_{x_{t_1}} \sigma_{x_{t_2}} + \rho_{c_{t_1},c_{t_2}}\sigma_{c_{t_1}} \sigma_{c_{t_2}} + \epsilon 
\label{equ:covdec}
\end{equation}
where ${\rho}_{x_{t_1},x_{t_2}}$ is the error autocorrelation in the original ensemble,
$\rho_{c_{t_1},c_{t_2}}$  the autocorrelation of the corrections,
$\sigma_{x_t}$ and $\sigma_{c_t}$ the standard deviation of the original ensemble and the standard deviation of the correction at lead time $t$, respectively.
The term $\epsilon$ corresponds to the estimated covariances of $x$ and $c$,
and is considered as negligible assuming that the original forecast and the corrections are drawn from two independent random processes.

Furthermore, the stationarity assumption of d-ECC implies that  
the correlation $\rho_{\tilde{x}_{t_1},\tilde{x}_{t_2}}$  
can also be estimated from past error statistics:
\begin{equation}
\rho_{\tilde{x}_{t_1},\tilde{x}_{t_2}}  = \mathbb{E}[{\hat{\rho}}_{e_{t_1},e_{t_2}}]
\end{equation}
where the notation $\hat{\rho}_{e_{t_1},e_{t_2}}$ refers to the elements of the estimated correlation matrix $\boldsymbol{\hat{R_e}}$.
The stationarity assumption takes effect in the transformation step of d-ECC (Eq.~\ref{eq:biascor}) which modifies the correlation of the scenario corrections  $\rho_{c_{t_1},c_{t_2}}$ and pushes it towards the estimated correlation $\hat{\rho}_{e_{t_1},e_{t_2}}$. 
In other words, the transformation affects $\rho_{c_{t_1},c_{t_2}}\sigma_{c_{t_1}} \sigma_{c_{t_2}}$ (second term in Eq.~\ref{equ:covdec}). We expect d-ECC to have a relevant impact if $\rho_{c_{t_1},c_{t_2}}\sigma_{c_{t_1}} \sigma_{c_{t_2}}$ dominates the sum in Eq.~\eqref{equ:covdec}. 
Typically, this is the case when the spread $\sigma_{x_t}$ of the original ensemble is small compared to the spread $\sigma_{\tilde{x}_t}$ after calibration. 
In a previous statement, we already noted that d-ECC takes effect if the calibration \emph{corrects} the spread. Regarding Eq.~\eqref{equ:covdec}, we can refine the statement and argue that d-ECC especially takes effect if the calibration \emph{increases} the spread. 
  
\begin{figure}
\centering
\includegraphics[width=7cm]{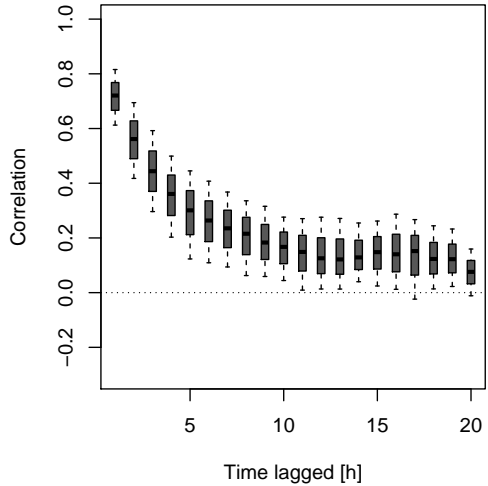}
\caption{
Temporal lagged correlation coefficients summarizing the error correlation matrix $\boldsymbol{\hat{R}}_e$ used in the d-ECC approach. The boxplots indicate the variability within the 3 month calibration period. 
}
\label{fig:gamma}
\end{figure}

Another important aspect of d-ECC is the estimation of the correlation matrix~$\boldsymbol{\hat{R}_e}$. By means of this matrix, the assumption of error autocorrelation is checked and adjusted. The matrix is estimated from the training datasets used for calibration at the different lead times. Based on the dataset described in Section \ref{sec:data}, Figure \ref{fig:gamma} shows the lagged correlation of the forecast error derived from $\boldsymbol{\hat{R}_e}$. The correlation is decreasing as a function of the time lag, reaching near zero values for lags greater than 10 hours. However, for short and very short time lags, the correlation is high and stable over the rolling training datasets. In particular, focusing on a time lag of 1 hour, the correlation ranges between 60\% and 80\%. 
The correlation variability  shown in Figure  \ref{fig:gamma} is estimated over a 3 month period. Similar results are obtained when checking the variability of the correlation  within each training dataset (not shown). The exhibited low variability indicates that the temporal correlation of the forecast error is not flow dependent. As a consequence, d-ECC can be seen as a "universal" approach that does not suffer restriction related to the forecasted weather situation. 


Considering again our case study, the scenarios generated with d-ECC based on the COSMO-DE-EPS forecasts are shown in Figure \ref{fig:exSC}(c).
The d-ECC derived scenarios are smoother and subjectively more realistic than the ones derived with ECC in Figure \ref{fig:exSC}(b). In Figure \ref{fig:excECC}, focusing on a single scenario, it is highlighted that the difference between the original and the d-ECC time trajectories varies gradually from one time interval to the next one while abrupt transitions occur in the case of the ECC scenario, as in this example between hours 15 and 17. 

Note that d-ECC does not give the same result as a simple smoothing of the calibrated scenarios~$\boldsymbol{\tilde{x}}$. Smoothing in time would modify the values $\boldsymbol{q}$ of the calibrated ensemble and possibly deteriorate its reliability. Instead, d-ECC affects the time variability of the scenarios by constructing a template (Eq.~\ref{equ:defz}) based on $\boldsymbol{\breve{x}}$ (Eq.~\ref{equ:kecc2}) while preserving the calibrated values $\boldsymbol{q}$.    

The discussion and illustration of d-ECC could certainly be extended by idealized studies and a rigorous mathematical framework. This would be welcomed as further research and would add further evidence to the expected behavior of d-ECC.  

\section{Verification methods}
\label{sec:verifmethod}

\subsection{Multivariate scores}
Verification of scenarios is first performed assessing the multivariate aspect of the forecast by means of adequate scores. The scores are applied focusing on scenarios in the form of time series. Considering an  ensemble with $N_e$ scenarios $\boldsymbol{x}^{(n)}$ with $n \in \left\{1,...,N_e \right\}$ and an observed scenario $\boldsymbol{y}$, the energy score \citep[$ES$;][]{gneiting2008} is defined as:
\begin{equation}
ES = \frac{1}{N_e} \sum_{n=1}^{N_e} \lVert \boldsymbol{y}-\boldsymbol{x}^{(n)} \rVert - \frac{1}{2N_e^2} \sum_{m=1}^{N_e}\sum_{p=1}^{N_e} \lVert \boldsymbol{x}^{(m)}-\boldsymbol{x}^{(p)} \rVert 
\label{equ:es}
\end{equation}
where $\|.\|$ represents the Euclidean norm.
ES is a generalization of the $CRPS$ to the multivariate case.

$ES$ suffers from a lack of sensitivity to misrepresentation of correlation structures \citep{pinsontastu2013}. We consider therefore additionally the p-variogram score \citep[$pVS$;][]{scheu2015}, which has better discriminative property in this respect. Based on the geostatistical concept of variogram, $pVS$ is defined as:
\begin{equation}
pVS = \sum_{i\ne j} \omega_{ij} \left( \mid y_i-y_j \mid^p 
- \frac{1}{N_e} \sum_{n=1}^{N_e} \mid x_i^{(n)}-x_j^{(n)} \mid^p \right)^2
\label{equ:pv}
\end{equation}
with $p$ the order of the variogram and  where $ \omega_{ij} $ are weights and the indices $i$ and $j$ indicate the $i$-th and the $j$-th components of the marked vectors, respectively. 
In order to focus on rapid changes in wind speed, the weights $ \omega_{ij} $ are chosen  proportional to the inverse square distance in time such:
\begin{equation}
 \omega_{ij}=\dfrac{1}{(i-j)^2},\quad i\ne j,
\label{equ:omega}
\end{equation}
since $i$ and $j$ are here forecast lead time indices.

\subsection{Multivariate rank histograms}
The multivariate aspect of the forecast is in a second step assessed by means of rank histograms applied to multi-dimensional fields \citep{thorar2014}. Two variants of the multivariate rank histogram are applied: the averaged rank histogram ($ARH$) and the band depth rank histogram ($BDRH$).  The difference of the two approaches lies in the way to defined pre-ranks from  multivariate forecasts.  $ARH$ considers the averaged rank over the multivariate aspect while $BDRH$ assesses the  centrality of the observation within the ensemble based on the concept of functional band depth.  

The interpretation of $ARH$ is the same as the interpretation of a univariate rank histogram:  $\cup$-shaped, $\cap$-shaped,  and flat rank histograms are interpreted as   underdispersiveness, overdispersiveness,  and  calibration of the underlying ensemble forecasts, respectively. The interpretation of $BDRH$ is different: a $\cup$-shape is associated to a lack of correlation, a $\cap$-shape to a too high correlation in the ensemble,  a skewed rank histogram to bias or dispersion errors and a flat rank histogram to calibrated forecasts.

\subsection{Product oriented verification }
Besides multivariate verification of time series scenarios, the forecasts are assessed in a product oriented framework. This type of scenario verification follows the spirit of the event oriented verification framework proposed by \cite{pins12}. Probabilistic forecasts that require time trajectories are provided and assessed by means of well-established univariate probabilistic scores.

Two types of products derived from forecasted scenarios are here under focus. The first one is defined as the mean wind speed over a day (here, a day is limited to the 21 hour forecast horizon). The second product is defined as the maximal upward wind ramp over a day, a wind ramp being defined as the difference between two consecutive forecast intervals.
For both products, 20 forecasts are derived from the 20 scenarios at each station and each verification day.

The performances of the ensemble forecasts for the two types of products are evaluated by means of the $CRPS$. The $CRPS$ is the generalization of the mean absolute error to predictive distributions \citep{gneiting2008}, and can be seen as the integral of the Brier score \citep[$BS$;][]{brier50} over all thresholds or the integral of the quantile score \citep[$QS$;][]{koenker78} over all probability levels. Considering an ensemble forecast, the $CRPS$ can be calculated as a weighted sum of $QS$  applied to the sorted ensemble members \citep{broecker012}. For a deeper insight in the forecast performance in terms of attributes, the $CRPS$ is decomposed following the same approach \citep{zbbmausam}: the $CRPS$ reliability and resolution components are calculated as weighted sums of the reliability and resolution components of the $QS$ at the probability levels defined by the ensemble size (see Eq.~\ref{equ:tau}), respectively. Formally, we write: 
\begin{equation}
CRPS_{reliability} = \frac{2}{N_e} \sum_{n=1}^{N_e} QS^{(\tau_n)}_{reliability}
\label{equ:crpsc1}
\end{equation}
\begin{equation}
CRPS_{resolution} = \frac{2}{N_e} \sum_{n=1}^{N_e} QS^{(\tau_n)}_{resolution}
\label{equ:crpsc1}
\end{equation}
where $QS^{(\tau_n)}_{reliability}$ and $QS^{(\tau_n)}_{resolution}$  are the reliability and resolution components of the $QS$ applied to the quantile forecasts at probability level $\tau_n$, respectively. The $QS$ decomposition is performed following \cite{Bentzien2014}. The $CRPS_{reliability}$ is negatively oriented (the lower the better) while the $CRPS_{resolution}$ is positively oriented (the higher the better).

\subsection{Bootstrapping}
The statistical significance of the results are tested applying a block-bootstrap approach. Boots\-trapping is a resampling technique which provides an estimation of the  statistical consistency  and is commonly applied to meteorological datasets \citep{efrontib86}.

A block-bootstrap approach is applied in the following which consists in defining a block as a single day of the verification period \citep{hamill99}. Each day is considered as a separate block of fully independent data. The verification process is repeated 500 times using each time a random sample with replacement of the 92 verification days (March, April,  May, 2013). The derived score distributions illustrate consequently the variability of the performance measures over the verification period and not between locations.  Boxplots are used to represent the distributions of the performance measures, where the quantile of the distributions at probability levels 5\%, 25\%, 50\%, 75 \% and 95\% are highlighted. 

\section{ Results and discussion}
\label{sec:results}

\begin{figure}
\centering
\includegraphics[width=7cm]{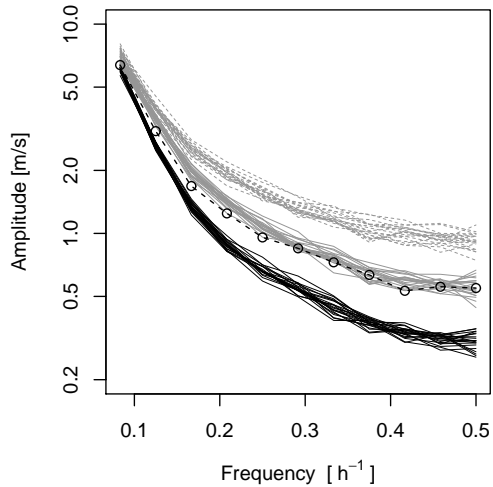}
\caption{
  Spectral analysis of the scenarios from the raw ensemble (black lines), of the scenarios derived with ECC (dashed grey lines) and with d-ECC (grey lines). Each line corresponds to one scenario among the 20.  The spectrum of the observed time series is represented by the dashed dotted line. 
}
\label{fig:spectr}
\end{figure}

Before applying the verification methods introduced in the previous section, 
we propose to explore statistically the time series variability by means of a spectral analysis, an analysis of the time series in the frequency domain.
Such an analysis is useful in order to describe statistical properties of the scenarios but has also direct implications for user's applications \citep[see below;][]{vincent2010}.
A Fourier transformation is applied to each forecasted and observed scenario and the contributions of the oscillations at various frequencies to the scenario variance examined \citep{wilksv1}. In Figure \ref{fig:spectr}, the mean amplitude of the forecast and observation time series over all stations and verification days is plotted as a function of their frequency components.

As already suggested by the case study, this analysis confirms that the ECC considerably increases the variability of the time trajectories with respect to the original ensemble, in particular at high frequencies. ECC scenario fluctuations are also much larger than the observed ones. Indeed, the amplitude is on average about two times larger at high frequencies in ECC time series than in the observed ones which explains the visual impression that ECC scenarios are unrealistic. Conversely, scenarios derived with the new copula approach do not exhibit such features. While the original ensemble shows a deficit of variability with respect to the observations, the d-ECC approach allows improving this aspect of the forecast. This first result, showing that d-ECC scenarios have a similar mean  spectrum as the observation one, is complemented with an objective assessment of the forecasted scenarios based on probabilistic verification measures.

\begin{figure*}
\centering
\includegraphics[width=14cm]{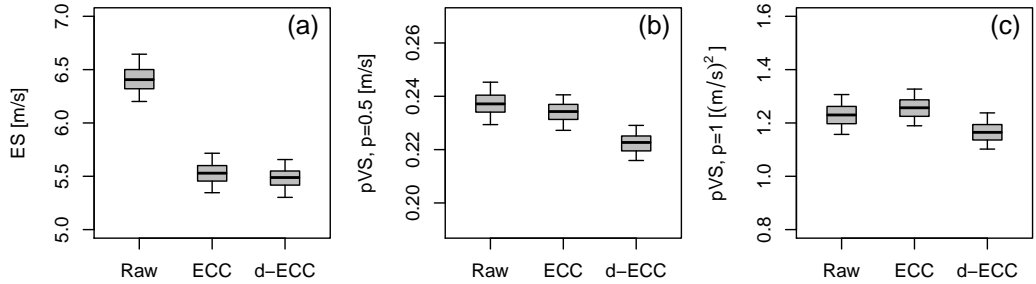}
\caption{
 Multivariate scores of time series: energy score (a) and p-variogram score for $p=0.5$ (b) and $p=1$ (c) in the form of box plots drawn from the application of a 500-block bootstrapping. 
}
\label{fig:res_ss}
\end{figure*}

Figure \ref{fig:res_ss} shows the performance of the forecasted time trajectories by means of multivariate scores. The post-processed scenarios perform significantly better than the raw members in terms of $ES$ (Figure \ref{fig:res_ss}(a)). In terms of $pVS$, the d-ECC scenarios are better than the ECC ones and significantly better than the raw ones when $p=0.5$ (Figure \ref{fig:res_ss}(b)). For higher orders of the variogram (here $p=1$, Figure \ref{fig:res_ss}(c)), the forecast improvement after post-processing is still clear when using d-ECC while the ECC results are 
slightly worse than 
the ones of the original forecasts.

\begin{figure*}
\centering
\includegraphics[width=14cm]{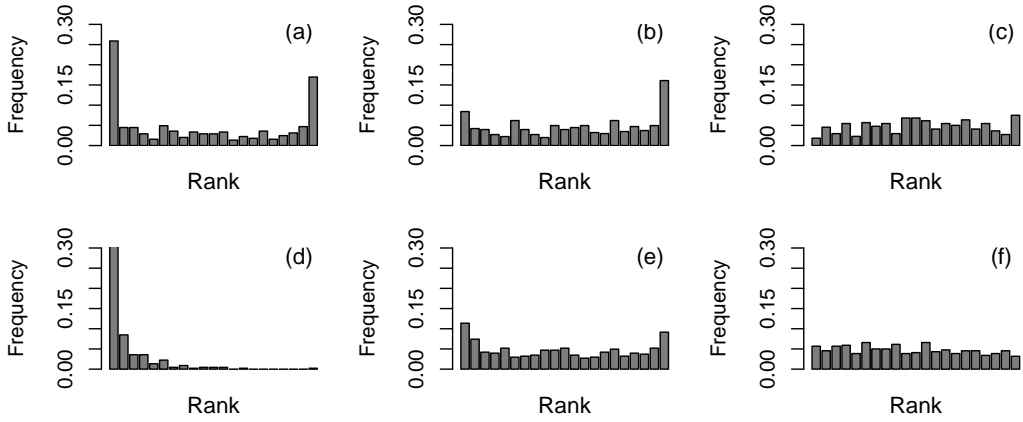}
\caption{
Multivariate rank histograms: (a,b,c) average rank histograms and (d,e,f) band depth rank histograms for time series from the raw ensemble (a,d) and derived with  ECC (b,e) and d-ECC (c,f).
}
\label{fig:res_rh}
\end{figure*}

Figure \ref{fig:res_rh} depicts the results in terms of multivariate rank histograms, $ARH$ (upper panel)  and $BDRH$ (lower panel).  The raw ensemble shows clear reliability deficiencies (Figures  \ref{fig:res_rh}(a) and \ref{fig:res_rh}(d)) which motivated the use of post-processing techniques.  Forecasts derived with ECC show still underdispersiveness but also too little correlation (Figures  \ref{fig:res_rh}(b) and \ref{fig:res_rh}(e)) while forecasts derived with d-ECC are better calibrated according to the rank histograms in Figures  \ref{fig:res_rh}(c) and  \ref{fig:res_rh}(f).
Indeed, both plots indicate good reliability of the d-ECC derived scenarios. 

\begin{figure*}
\centering
\includegraphics[width=14cm]{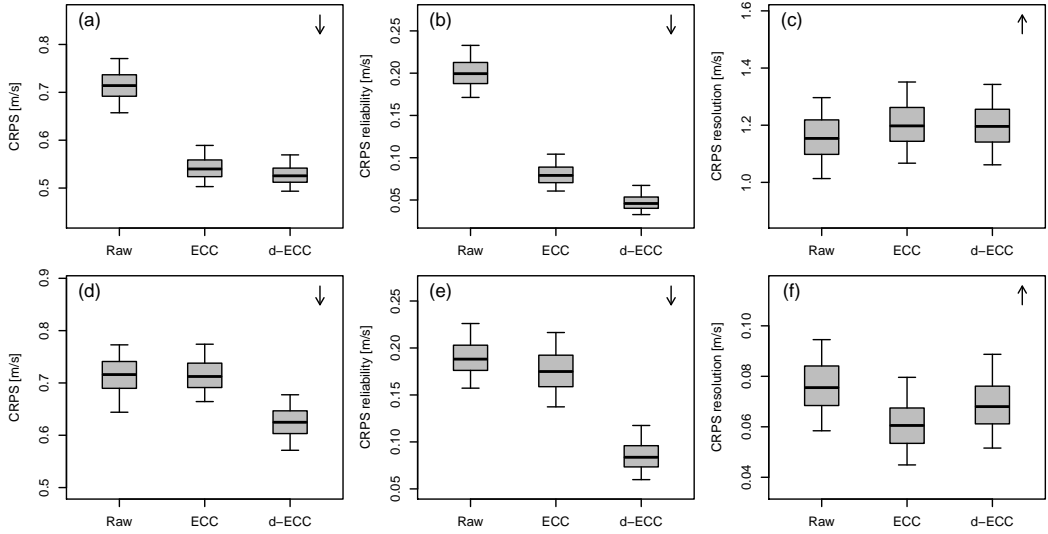}
\caption{
  Product oriented verification of scenarios: 
  (a,b,c) daily  means at station,  
  (d,e,f) maximal upward ramps  within a day at station. 
  Results are shown in terms of $CRPS$ (a,d), $CRPS$ reliability component (b,e) and $CRPS$ resolution component (c,f). The box plots  indicate confidence intervals estimated with block bootstrapping. The arrows in the right corners indicate whether the performance measure is positively or negatively oriented. 
}
\label{fig:res_prod}
\end{figure*}

Figure \ref{fig:res_prod} focuses on two products drawn from the time series forecasts: the daily mean wind speed (upper panel) and the daily maximal upward ramp (lower panel). The performances are assessed in terms of $CRPS$, $CRPS$ reliability and $CRPS$ resolution, from left to right, respectively. Looking at the results in terms of $CRPS$, we  note the high similarity of Figures  \ref{fig:res_prod}(a) and \ref{fig:res_prod}(d) with Figures \ref{fig:res_ss}(a) and \ref{fig:res_ss}(c), respectively. As for the $ES$, post-processing significantly improves the forecasts of the daily mean product. As for $pVS$ with $p=1$, d-ECC improves the ramp product with respect to the original one while ECC does not generate improved products. 
The $CRPS$ decomposition allows detailing the origin of these performances. We see in Figures \ref{fig:res_prod}(b) and \ref{fig:res_prod}(e) that the $CRPS$ results are mainly explained by the impact of the post-processing on the $CRPS$ reliability components. However, focusing on the results in terms of $CRPS$ resolution in Figures \ref{fig:res_prod}(c) and \ref{fig:res_prod}(f), we note that the resolution of the original and d-ECC products are comparable while ECC deteriorates the resolution of the ramp product with respect to the original one.    

Those verification results are interpreted as follows. Calibration corrects for the mean of the ensemble forecast and this is reflected, after the derivation of scenarios, by an improvement of the $ES$ and daily mean product skill. Calibration also corrects for spread deficiencies increasing the variability of the ensemble forecasts. This increase of spread associated with a preservation of the rank structure of the original ensemble, as it is the case in the ECC approach, enlarges indiscriminately the temporal variability of the forecasts and leads to 
a slight deterioration of the $pVS$ and ramp product results.

The d-ECC approach provides scenarios with a temporal variability comparable to the one of the observation. In that case, the benefit of the calibration step in terms of reliability (at single forecast lead times) persists at the multivariate level (looking at time trajectories) after the reconstruction of scenarios with d-ECC. The multivariate reliability, or the reliability of derived products, 
is significantly improved after post-processing, though not perfect for specific derived products. Moreover, d-ECC scenarios perform as well as the original ensemble forecast in terms of resolution. 
So, unlike ECC, d-ECC is able  to generate reliable scenarios with a level of resolution  that is not deteriorated with respect to the original ensemble forecasts.

\section{Conclusion and outlook}
\label{sec:conc}

A new empirical copula approach is proposed for the post-processing of calibrated ensemble forecasts. The so-called dual ensemble copula coupling approach is introduced with a focus on temporal structures of wind forecasts. The new scheme includes a temporal component in the ECC approach accounting for the error autocorrelation of the ensemble members. The estimation of the correlation structure in the error based on past data allows adjusting the dependence structure in the original ensemble. 

Based on COSMO-DE-EPS forecasts, the scenarios derived by d-ECC prove to be qualitatively realistic and quantitatively of superior quality. Post-processing of wind speed combining EMOS and d-ECC improves the forecasts in many aspects. In comparison to ECC, d-ECC drastically improves the quality of the derived scenarios. Applications that require temporal trajectories will fully benefit of the new approach in that case. As for any post-processing technique, the benefit of the new copula approach  can be weakened by improving the representation of the forecast uncertainty with more efficient member generation techniques and/or by improving the calibration procedure correcting for conditional biases. 
Meanwhile, at low additional complexity and computational costs, d-ECC can be considered as a valuable alternative to the standard ECC for the generation of consistent scenarios.

Though only the temporal aspect has been investigated in this study, the dual ensemble copula approach could be generalized to any multivariate setting. 
Further research is however required for the application of d-ECC at scales that are unresolved by the observations. For example, geostatistical tools could be applied for the description of the autocorrelation error structure at the model grid level. Moreover, the mathematical interpretation of the d-ECC scheme developed here would benefit from further theoretical investigations based on idealized case studies.  


\section*{Acknowledgments}
This work has been done within the framework of the EWeLiNE  project (\textit{Erstellung innovativer Wetter- und Leistungsprognosemodelle 
f\"ur die Netzintegration wetterabh\"angiger Energietr\"ager}) funded by the German Federal Ministry for Economic Affairs and Energy. The authors acknowledge the Department of Wind Energy of the Technical University of Denmark (DTU), the German Wind Energy Institute (DEWI GmbH), DNV GL, the Meteorological Institute (MI) of University of Hamburg and the Karlsruhe Institute of Technology (KIT) for providing wind measurements at stations Risoe, FINO1 and FINO3, FINO2, Hamburg and Lindenberg, and Karlsruhe, respectively.  
The authors are also grateful to Tilmann Gneiting and two anonymous reviewers  for helpful and accurate comments on a previous version of this manuscript.

\bibliographystyle{wileyqj}

\end{document}